\documentclass[prl,twocolumn,amsmath,amssymb,aps,superscriptaddress]{revtex4}
\usepackage{graphicx}
\usepackage{amsmath,amssymb,mathrsfs}

\def\Tr{\operatorname{Tr}} 
\def\>{\rangle}\def\<{\langle} \def\sH{\mathscr{H}}
   
\def\mE{\mathcal{E}}

\def\instr{\mathscr{M}} \def\info{\iota(\rho^Q,\mathscr{M}^Q)}

\def\dist{\delta(\rho^Q,\mathscr{M}^Q)} \def\out{\mathcal{X}}
\def\mR{\mathcal{R}} 
\def\id{\operatorname{id}} \def\noise{\Delta(\rho^Q,\instr^Q)}

\newtheorem{theo}{Theorem}

\begin{document}

\title{\bf Global information balance in quantum measurements}

\author{Francesco Buscemi} \email{buscemi@qci.jst.go.jp}
\affiliation{ERATO-SORST Quantum Computation and Information Project,
  Japan Science and Technology Agency}

\author{Masahito Hayashi} \affiliation{ERATO-SORST Quantum
  Computation and Information Project, Japan Science and Technology
  Agency} \affiliation{Graduate School of Information Sciences, Tohoku
  University}

\author{Micha\l{} Horodecki} \affiliation{Institute of Theoretical
  Physics and Astrophysics, University of Gda\'nsk, Poland}

\date{May 7, 2008}

\begin{abstract}
  We perform an information-theoretical analysis of quantum
  measurement processes and obtain the global information balance in
  quantum measurements, in the form of a closed chain equation for
  quantum mutual entropies. Our balance provides a tight and general
  entropic information-disturbance trade-off, and explains the
  physical mechanism underlying it. Finally, the single-outcome case,
  that is, the case of measurements with post-selection, is briefly
  discussed.
\end{abstract}

\maketitle

It is now well-known that, even if Heisenberg uncertainty relations do
not describe the disturbance caused on a quantum system by a quantum
measurement~\cite{ballentine,ozawa11,dariano}, Quantum Mechanics
\emph{does} indeed provide the existence of a monotonic
information-disturbance relation: the more useful information is
extracted from a quantum system, the more such a system is disturbed
by the measurement. This fact is apparent from general arguments (if
it were possible to gain information without causing disturbance, then
it would be possible to determine the wave-function of an arbitrary
system~\cite{dariano-yuen}), as well as from some explicitly derived
tradeoff relations obtained for some specific estimation
tasks~\cite{tradeoffs}. On the other hand, proposals of reversible
measurements have been reported~\cite{invert}, all of them implicitly
based on the existence of a tradeoff between information extraction
and probability of exact correction.

Despite the enormous relevance a universal relation between
information extraction and disturbance due to a quantum measurement
would have from both foundational and practical point of view, a
general approach to the problem of quantifying such a relation, quite
surprisingly, is still lacking. The main difficulty seems to be that
explicitly known tradeoff relations involve quantities (like e.~g. the
average error probability or the average output fidelity) which
strongly rely on the way the classical signals are encoded into
quantum states, i.e. on the structure of the input ensemble. For this
reason, the only known tradeoff curves cover some specific classes of
ensembles enjoying symmetry properties, making the derivation
possible~\cite{tradeoffs}.

The approach we propose in this Letter in order to overcome such a
specificity, is to work with genuinely quantum entities. More
explicitly, we will introduce a \emph{quantum information gain}, which
constitutes an upper bound to the information that the apparatus is
able to extract, independently of how this information is encoded, and
a \emph{quantum disturbance}, which is related to the possibility of
deterministically and coherently undoing the corresponding state
change. Both quantum information gain and quantum disturbance are
intimately related to previously known and independent notions:
Groenewold's information gain~\cite{groene} on one side, and channels
coherent information~\cite{schum-lloyd} on the other, the latter
applied by Maccone~\cite{macca} as a measure of disturbance, in the
first attempt to ``quantize'' tradeoff relations. However, both of the
quantities, as they were originally introduced, are not applicable to
the most general situation. The definitions we introduce here, not
only constitute a proper reformulation of these latter, but also allow
us to elegantly link these (previously independent) quantities by
using the chain rule for quantum mutual information only, thus
establishing a closed information balance in quantum
measurements. Such a balance provides, as a built-in feature, a tight
and general entropic information-disturbance tradeoff relation. The
same approach will be shown to be straightforwardly applicable also to
the case of measurements with post-selection.

{\it Quantum instruments.}--- A general measurement process $\instr^Q$
on the input system $Q$, described by the input density matrix
$\rho^Q$ on the (finite-dimensional) Hilbert space $\sH^Q$, can be
described as a collection of classical outcomes $\out:=\{m\}$,
together with a set of completely positive (CP) maps
$\{\mE_m^Q\}_{m\in\out}$~\cite{kraus}, such that, when the outcome $m$
is observed with probability $p(m):=\Tr[\mE_m^Q(\rho^Q)]$,
$\sum_mp(m)=1$, the corresponding \emph{a posteriori} state
$\rho^{Q'}_m:=\mE_m^Q(\rho^Q)/p(m)$ is output by the apparatus. This
is the CP quantum instruments formalism introduced by
Ozawa~\cite{ozawa-instr} (\footnote{We put a prime on the output
  system to include situations where the input physical system ($Q$)
  gets transformed into something else ($Q'$). This is the case of
  \emph{demolishing} measurements: an outcome happens to occur if and
  only if there exists the corresponding \emph{a posteriori}
  state---maybe carried by a different quantum system.}). With a
little abuse of notation, we can think that the action of the
measurement $\instr^Q$ on $\rho^Q$ is given in average by the mapping
\begin{equation}\label{eq:average_instr}
\begin{split}
\instr^Q(\rho^Q)&:=\sum_mp(m)\rho^{Q'}_m\otimes m^\out\\
&:=\Theta^{Q'\out},
\end{split}
\end{equation}
where $\{|m^\out\>\}_m$ is a set of orthonormal (hence perfectly
distinguishable) vectors on the classical register space $\out$ of
outcomes. If the outcomes are discarded before being read out, that
is, if $\Theta^{Q'\out}$ in Eq.~(\ref{eq:average_instr}) is traced
over $\out$, then the resulting average map $\mE^Q:=\sum_m\mE^Q_m$ is
a channel, i.~e. a CP trace-preserving (TP) map: quantum instruments
contain quantum channels as a special case. If, on the other hand, we
are not interested in the \emph{a posteriori} states but only in the
outcomes probability distribution $\vec p(m)$ (that is equivalent to
tracing $\Theta^{Q'\out}$ over $Q'$), then the resulting average map
$\rho^Q\mapsto\vec p(m)$ is described by a Positive Operator Valued
Measure (POVM), namely, a set of positive operators
$\{P^Q_m\}_{m\in\out}$, $\sum_mP^Q_m=\openone^Q$, such that
$p(m)=\Tr[\rho^Q P^Q_m]$: quantum instruments contain POVMs as a
special case.

We now exploit a very useful representation theorem for CP quantum
instruments~\cite{ozawa-instr}: it states that whatever quantum
measurement can be modeled as an \emph{indirect measurement}, in which
the input system first interacts with an \emph{apparatus} (or probe)
$A$, initialized in a fixed pure state $\phi^A$, through a suitable
unitary interaction $U^{QA}:QA\to Q'A'\simeq QA$; subsequently, a
particular measurement $\instr^{A'}$, depending also on $U^{QA}$, is
performed on the apparatus. In addition, by introducing a third
\emph{reference} system $R$ purifying the input state as $\Psi^{RQ}$,
$\Tr_R[\Psi^{RQ}]=\rho^Q$, we are in the situation schematically
represented as in FIG.~1: right after the unitary interaction
$U^{QA}$, the global tripartite state is $|\Upsilon^{RQ'A'}\>
:=(\openone^R\otimes U^{QA}) (|\Psi^{RQ}\>\otimes|\phi^A\>)$, and the
measurement on the apparatus can be chosen such that~\footnote{From
  Ref.~\cite{ozawa-instr}, $U^{QA}$ and $\instr^{A'}$ can be chosen
  such that $\mE^{A'}_m(\rho^{A'})=K_m\rho^{A'}K_m^\dag$, $\forall m$,
  with $\sum_mK_m^\dag K_m=\openone^{A'}$.}
\begin{equation}\label{eq:average_instr_pure}
\begin{split}
(\id^{RQ'}\otimes\instr^{A'})(\Upsilon^{RQ'A'})&:=\sum_mp(m)\Upsilon^{RQ'A''}_m\otimes m^\out\\
&:=\Theta^{RQ'A''\out},
\end{split}
\end{equation}
where $\Upsilon^{RQ'A''}_m$ are \emph{pure} states such that
$\Tr_{A''}[\Upsilon^{RQ'A''}_m] =(\id^R\otimes\mE^Q_m)(\Psi^{RQ})/p(m)
=:\rho^{RQ'}_m$ and $\Tr_{R}[\rho^{RQ'}_m]=\rho^{Q'}_m$, and $m^\out$
are the classical register states, as before. The above equation is
nothing but a particular extension of Eq.~(\ref{eq:average_instr}): in
fact, by tracing $\Theta^{RQ'A''\out}$ over $R$ and $A''$, one obtains
the state $\Theta^{Q'\out}$ in Eq.~(\ref{eq:average_instr}). (For this
reason, in the following, where no confusion arises, we will adopt the
convention that to omit indices in the exponent of a multipartite
state means to trace over the omitted indices.) Even though it is a
simple rewriting, Eq.~(\ref{eq:average_instr_pure}) will turn out to
be very useful for our analysis, in that it gives a deeper insight in
understanding the overall information balance.

\begin{figure}[h]
  \includegraphics[width=8cm]{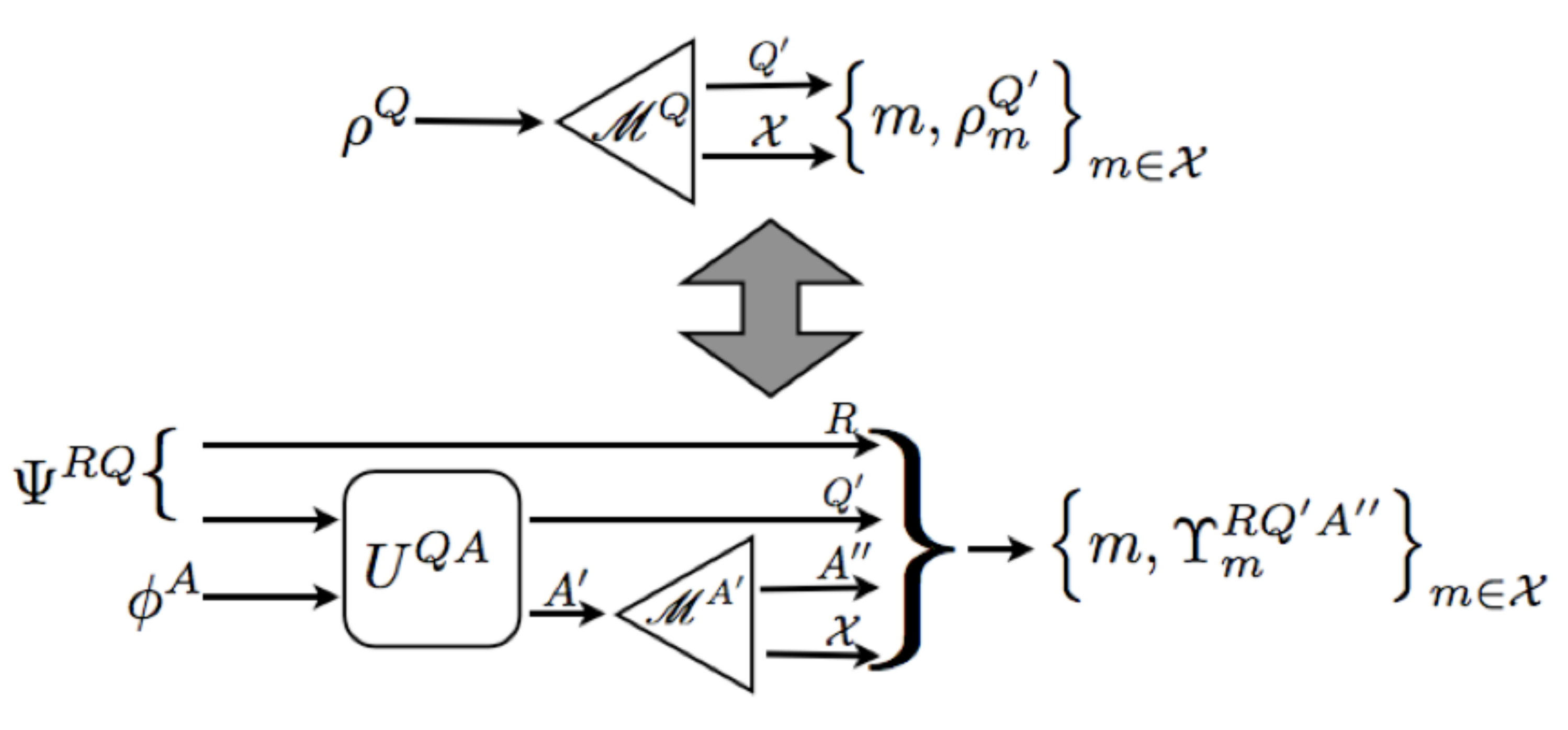}
  \caption{The action of a quantum measurement $\instr^Q$ on the input
    state $\rho^Q$ (top), can always be extended as a tripartite
    indirect measurement (bottom), where the apparatus $A$, after
    having properly interacted with the input system through $U^{QA}$,
    undergoes the measurement $\instr^{A'}$. The conditional output
    pure states $\{\Upsilon^{RQ'A''}_m\}_m$ are such that
    $\Tr_{RA''}[\Upsilon^{RQ'A''}_m]=\rho^{Q'}_m$, $\forall m$. See
    Eqs.~(\ref{eq:average_instr}) and~(\ref{eq:average_instr_pure}) in
    the text.}
  \label{fig:1}
\end{figure}

{\it Quantum information gain.}--- Having in mind
Eq.~(\ref{eq:average_instr_pure}), we define the (quantum)
\emph{information gain} $\iota(\rho^Q,\instr^Q)$ of the measurement
$\instr^Q$ on the input state $\rho^Q$ as
\begin{equation}\label{eq:info}
\iota(\rho^Q,\instr^Q):=I^{R:\out}(\Theta^{R\out}),
\end{equation}
where $I^{A:B}(\sigma^{AB}) :=S(\sigma^A) +S(\sigma^B)
-S(\sigma^{AB})$ is the usual \emph{quantum mutual
  information}~\cite{mutual_info}. Due to the particular form of
$\Theta^{RQ'A''\out}$, it is possible to rewrite such a quantity as
$\iota(\rho^Q,\instr^Q)=S(\rho^{R})-\sum_mp(m)S(\rho^{R}_m)$, for
$\rho^R_m:=\Tr_{Q'}[\rho^{RQ'}_m]$. In other words,
$\iota(\rho^Q,\instr^Q)$ is the $\chi$-quantity~\cite{chi} of the
ensemble induced on the system $R$ by the measurement $\instr^Q$. In
communication theory in fact, the information gain is usually better
understood as being about the remote system $R$, while $Q$, correlated
with $R$, represents just the carrier that is measured. Nonetheless,
it is a crucial point, for what follows, that the information
gain~(\ref{eq:info}) only depends on the input state $\rho^Q$ and on
the measurement $\instr^Q$ performed onto it, regardless of the
particular extension constructed in
Eq.~(\ref{eq:average_instr_pure}),~\footnote{This further justifies
  the explicit appearance of $R$---a formally defined, hence seemingly
  ``unphysical'', system---into our definitions of quantum information
  gain~(\ref{eq:info}) and, later, of quantum
  disturbance~(\ref{eq:dist}).}. Indeed $\info$ depends only on the
input state $\rho^Q$ and on the POVM $\{P^Q_m\}_{m\in\out}$ induced by
$\instr^Q$, regardless of the particular state reduction maps
$\{\mE^Q_m\}_{m\in\out}$ and of the explicit form of the \emph{a
  posteriori} states $\{\rho^{Q'}_m\}_{m\in\out}$.

Holevo's upper bound~\cite{holevo} on the accessible information
provides a clear interpretation of our definition of information gain:
$\iota(\rho^Q,\instr^Q)$ \emph{is the Holevo bound to the amount of
  classical information which can be reliably extracted by the
  measurement} $\instr^Q$ \emph{from the input state} $\rho^Q$. In
fact, consider whatever classical alphabet $X:=\{x\}$, however encoded
on the input state as $\rho^Q=\sum_x\rho^Q_x$: in this case, the joint
input-output probability distribution is given by
$p(x,m)=\Tr[\mE_m^Q(\rho^Q_x)]$. On the other hand, every such an
encoding can be (formally) seen as induced by the measurement of a
suitable POVM over the reference system $R$, in formula,
$\rho^Q_x=\Tr_{R}[(P^R_x\otimes\openone^Q) \Psi^{RQ}]$, for some POVM
$\{P^R_x\}_{x\in X}$. This means that we can also write
$p(x,m)=\Tr[\rho^{R}_mP^{R}_x]$, implicitly considering a ``dual''
situation, in which the encoded input is $m$ and the decoded letter is
$x$. It is clear then, that the classical mutual information
$I(X:\out)$---which is a symmetric function of its arguments---between
the alphabet $X$ and the indices in $\out$ is upper bounded by the
$\chi$-quantity of the ensemble $\{p(m),\rho^R_m\}_{m\in\out}$ which
exactly corresponds to $\iota(\rho^Q,\instr^Q)$. In formula:
$I(X:\out)\le\iota(\rho^Q,\instr^Q)$.

It is interesting here to compare our definition of information gain
to the one dating back to Groenewold~\cite{groene} (and which, by the
way, was never put in relation with whatsoever notion of
disturbance). He defined the information gain for von Neumann-L\"uders
measurements to be equal to
$\iota_G(\rho^Q,\instr^Q):=S(\rho^Q)-\sum_mp(m)S(\rho^{Q'}_m)$,
conjecturing its positivity. Later Ozawa~\cite{ozawa} generalized
Groenewold's definition to take into account all possible measurements
and characterized those with $\iota_G\ge 0$, explicitly pointing out
that the general quantum instruments formalism commonly allows
situations where $\iota_G<0$. This feature, making the interpretation
of $\iota_G$ as an information gain problematic, comes from the fact
that Groenewold-Ozawa definition, contrarily to ours, explicitly
depends on the particular \emph{a posteriori} states
$\{\rho^{Q'}_m\}_{m\in\out}$. Nonetheless, there are situations (to be
shown in the following) where
$\iota_G(\rho^Q,\instr^Q)=\iota(\rho^Q,\instr^Q)$. Incidentally, our
definition of information gain~(\ref{eq:info}) always returns the same
numerical value of Winter's ``intrinsic information'' of a
POVM~\cite{Winter}, thus gaining an operational interpretation, and of
Hall's ``dual upper bound'' on accessible information~\cite{Hall},
even if their definitions slightly differ from ours.

{\it Quantum disturbance.}--- As we anticipated in the introduction,
our notion of disturbance is closely related to that of coherent
information. A first step in this direction is due to
Maccone~\cite{macca}, who however used a different definition, not
suitable for the case of general quantum measurements. We define the
(quantum) \emph{disturbance} $\dist$ caused by the measurement
$\instr^Q$ on the input state $\rho^Q$ as
\begin{equation}\label{eq:dist2}
\dist:=S(\rho^Q)-I_c^{R\to Q'\out}(\Theta^{RQ'\out}),
\end{equation}
where $I_c^{A\to B}(\sigma^{AB}):=S(\sigma^B)-S(\sigma^{AB})$ is the
so-called \emph{coherent information}~\cite{schum-lloyd}. In the
following we will show the reason why the quantity $\dist$ can be
understood as the disturbance.

Given a quantum channel $\mE^Q$, from $Q$ to $Q'$, acting on the input
state $\rho^Q$, it is known that the coherent information $I_c^{R\to
  Q'}((\id^R\otimes\mE^Q) (\Psi^{RQ}))$ plays a central role in
quantifying how well the channel preserves quantum coherence. In fact,
coherent information turns out to be intimately related to the
possibility of constructing a recovering operation $\mR^{Q'}$, from
$Q'$ to $Q$, correcting the action of $\mE^Q$: the closer the coherent
information is to its maximum value $S(\rho^Q)$, the closer (on the
support of $\rho^Q$) the corrected channel $\mR^{Q'}\circ\mE^Q$ is to
the ideal channel $\id^Q$. In particular, in
Ref.~\cite{schumacher-westmoreland} it is proved that whenever
$S(\rho^Q)-I_c^{R\to Q'}((\id^R\otimes\mE^Q) (\Psi^{RQ}))\le\epsilon$,
then it is possible to explicitly construct a correcting channel
(generally depending also on $\rho^Q$, but for sake of clarity of
notation, we will drop such dependence, leaving it understood)
$\mR^{Q'}$ such that $F_e(\rho^Q, \mR^{Q'}\circ\mE^{Q})\ge
1-2\sqrt{\epsilon}$, where $F_e(\rho^Q,\mR^{Q'}
\circ\mE^{Q}):=\<\Psi^{RQ}| (\id^{R}\otimes\mR^{Q'}\circ\mE^Q)
(\Psi^{RQ})|\Psi^{RQ}\>$ is the \emph{entanglement
  fidelity}~\cite{schum} of the corrected channel
$\mR^{Q'}\circ\mE^{Q}$ with respect to the input state $\rho^Q$. The
value of $F_e(\rho^Q,\mR^{Q'} \circ\mE^Q)$ says how close is the
corrected channel $\mR^{Q'}\circ\mE^Q$ to the identity channel
$\id^{Q}$ on the support of $\rho^Q$. If such value is close to one,
it means not only that $\mR^{Q'}(\mE^Q(\rho^{Q}))$ is close to
$\rho^Q$, but also that quantum correlations between $Q$ and $R$ are
almost preserved.

The correction exploited in Ref.~\cite{schumacher-westmoreland} is
blind, in the sense that the channel $\mR^{Q'}$ is a fixed one and
works well on the average channel $\mE^{Q}$. In our setting, on the
contrary, the indices $m$ are by definition visible, in that they are
the outcomes of the measurement: this fact reflects the form of
Eq.~(\ref{eq:dist2}), where the output $Q'$ is considered
\emph{jointly} with the outcomes space $\out$. Then,
following~\cite{schumacher-westmoreland}, a fixed correcting channel
$Q'\out\to Q$ results in a family of correcting channels
$\mR_m^{Q'}:Q'\to Q$, depending on the measurement readout $m$. We
thus obtained the following
\begin{theo}[Approx. measurement correction]
  If $\dist\le\epsilon$, then there exists a family of recovering
  operations $\{\mR_m^{Q'}\}_{m\in\out}$ such that
\begin{equation*}
  F_e(\rho^Q,\sum_m\mR_m^{Q'}\circ\mE^Q_m)\ge 1-2\sqrt{\epsilon}.\qquad\square
\end{equation*}
\end{theo}
It is worth stressing that also the converse statement is true,
namely, an approximately reversible instrument is almost
undisturbing. In fact, as proved in Ref.~\cite{barnischu}, a sort of
quantum Fano inequality holds for every set of channels
$\{\mR_m^{Q'}\}_{m\in\out}$, in that $\dist\le
\operatorname{f}[1-F_e(\rho^Q,\sum_m\mR_m^{Q'}\circ\mE^Q_m)]$, where
$\operatorname{f}(x)$ is an appropriate positive, continuous,
monotonic increasing function such that $\operatorname{f}(0)=0$.

As we said, our definition of disturbance~(\ref{eq:dist2}) generalizes
the usual notion of coherent information loss for quantum channels,
which can be recovered from our formula~(\ref{eq:dist2}) by simply
tracing over the outcomes space $\out$, thus obtaining the quantity
$S(\rho^Q)-I_c^{R\to Q'}(\Theta^{RQ'})$, which, thanks to the
data-processing inequality, is always greater than or equal to
$S(\rho^Q)-I_c^{R\to Q'\out}(\Theta^{RQ'\out})$. In other words, when
discarding the outcomes (as done in Ref.~\cite{macca}), the
disturbance is higher, thus providing a too much loose tradeoff. The
importance of taking into account the measurement outcomes during the
correction is then clear~\cite{bus}.

{\it Global information balance.}--- Before proceeding, let us
explicitly calculate the disturbance~(\ref{eq:dist2}) for the state
$\Theta^{RQ'A''\out}$: because of the classical feature of $\out$, we
find
\begin{equation}\label{eq:dist}
\dist=I^{R:A''\out}(\Theta^{RA''\out}).
\end{equation}
Then, by using the chain rule for quantum mutual
information~\cite{hayashi}, valid for all tripartite states
$\sigma^{ABC}$, that is $I^{A:C}(\sigma^{AC}) +I^{A:B|C}(\sigma^{ABC})
=I^{A:BC}(\sigma^{ABC})$, where
$I^{A:B|C}(\sigma^{ABC}):=S(\sigma^{AC}) +S(\sigma^{BC})
-S(\sigma^{ABC}) -S(\sigma^{C})$ is the \emph{quantum conditional
  mutual information}, we can put together Eqs.~(\ref{eq:info})
and~(\ref{eq:dist}), thus obtaining the global balance of information
in a quantum measurement as
\begin{equation}\label{eq:balance}
\info+\noise=\dist.
\end{equation}
The positive quantity
\begin{equation}\label{eq:noise}
\begin{split}
\noise&:=I^{R:A''|\out}(\Theta^{RA''\out})\\
&=\sum_mp(m)I^{R:A''}(\rho^{RA''}_m),
\end{split}
\end{equation}
for $\rho^{RA''}_m:=\Tr_{Q'}[\Upsilon^{RQ'A''}_m]$, measures the
``missing information'' in terms of the hidden correlations between
the reference system and some inaccessible degrees of
freedom---internal degrees of freedom of the apparatus or
environmental degrees of freedom which interacted with the apparatus
during the measurement process---which cannot be controlled by the
experimenter~\footnote{Compare our information balance to other
  conservation-like relations, see e.~g., M.~Jakob and J.~A.~Bergou,
  preprint arXiv:quant-ph/0302075v1 (2003), and M.~Horodecki \emph{et
    al.}, Found.~Phys.~{\bf 35}, 2041 (2005).}. The existence of a
tradeoff between information gain and disturbance is then a direct
evidence of the appearance of such correlated hidden degrees of
freedom.

The quantity $\noise$ is null if and only if, for every outcome $m$,
the reference and the apparatus are in a factorized state, that is,
$\rho^{RA''}_m=\rho^R_m\otimes\rho^{A''}_m$, $\forall m$. This is the
case, for example, of the so-called ``single-Kraus'' or ``multiplicity
free'' instruments, for which every map $\mE^Q_m$ is represented by a
single contraction as $\mE^Q_m(\rho^Q)=E_m\rho^Q E_m^\dag$, with
$E_m^\dag E_m\le\openone^Q$. Hence, this kind of measurements maximize
the information gain for a fixed disturbance, or, equivalently,
minimize the disturbance for a fixed information gain: they are
optimal measurements---in a sense, noiseless---closely related to the
notion of ``clean measurements'' introduced in
Ref.~\cite{clean}. Single-Kraus measurements satisfy $\info=\dist$,
while, in general cases, the tradeoff $\info\le\dist$ holds. Moreover,
for single-Kraus measurements, Groenewold-Ozawa information gain
coincide with ours, namely,
$\iota_G(\rho^Q,\instr^Q)=\iota(\rho^Q,\instr^Q)$, as anticipated
before.

{\it Measurements with post-selection.}--- It is a remarkable
advantage of our approach, the fact that the analysis of the
single-outcome case is possible. The importance of such an analysis is
strongly motivated by D'Ariano in Ref.~\cite{dariano}. Let us define
the single-outcome versions of Eqs.~(\ref{eq:info}),~(\ref{eq:dist2}),
and~(\ref{eq:noise}) as $\iota_m(\rho^Q,\instr^Q)
:=S(\rho^R)-S(\rho^R_m)$, $\delta_m(\rho^Q,\instr^Q)
:=S(\rho^Q)-I_c^{R\to Q'}(\rho^{RQ'}_m)$, and
$\Delta_m(\rho^Q,\instr^Q) :=I^{R:A''}(\rho_m^{RA''})$. These three
quantities satisfy the analogous of Eq.~(\ref{eq:balance}), that is
$\iota_m(\rho^Q,\instr^Q) +\Delta_m(\rho^Q,\instr^Q)
=\delta_m(\rho^Q,\instr^Q)$. Notice now that, while
$\Delta_m(\rho^Q,\instr^Q)$ is always positive (since it is a quantum
mutual information), both $\iota_m(\rho^Q,\instr^Q)$ and
$\delta_m(\rho^Q,\instr^Q)$ can assume \emph{negative} values: while a
negative conditional information gain can well understood also in
classical information theory, a negative conditional disturbance
simply means that the entanglement between $R$ and $Q'$, conditionally
on a particular outcome, is higher than the original entanglement in
$|\Psi^{RQ}\>$.

{\it Acknowledgments.}--- We would like to thank A.~Barchielli,
G.~M.~D'Ariano, R.~Horodecki, M.~Ozawa, and M.~F.~Sacchi for
enlightening comments. F.~B. and M.~Ha. acknowledge Japan Science and
Technology Agency for support through the ERATO-SORST Quantum
Computation and Information Project.  M.~Ho. is supported by EC IP
SCALA. Part of this work was done while M.~Ho. was visiting
ERATO-SORST project.

\appendix


\begin{thebibliography}{99}
\bibitem{ballentine} L.~E.~Ballentine, Rev.~Mod.~Phys.~{\bf 42}, 358
  (1970).
\bibitem{ozawa11} M.~Ozawa, Phys.~Lett.~A {\bf 282}, 336 (2001);
  M.~Ozawa, Ann.~Phys.~{\bf 311}, 350 (2004).
\bibitem{dariano} G.~M.~D'Ariano, Fortschr. Phys. {\bf 51}, 318 (2003).
\bibitem{dariano-yuen} A.~Royer, Phys.~Rev.~Lett. {\bf 73}, 913 (1994),
  and \emph{ibid.}~{\bf 74}, 1040 (1995); G.~M.~D'Ariano and H.~P.~Yuen,
  \emph{ibid.}~{\bf 76} 2832 (1996).
\bibitem{tradeoffs} C.~A.~Fuchs and A.~Peres, Phys.~Rev.~A {\bf 53}, 2038
  (1996); K.~Banaszek, Phys.~Rev.~Lett. {\bf 86}, 1366 (2001);
  K.~Banaszek and I.~Devetak, Phys.~Rev.~A {\bf 64}, 052307 (2001);
  L.~Mi\v{s}ta Jr., J.~Fiur\'a\v{s}ek, and R.~Filip, \emph{ibid.}~{\bf
    72}, 012311 (2005); M.~F.~Sacchi, Phys.~Rev.~Lett.  {\bf 96}, 220502
  (2006); M.~G.~Genoni and M.~G.~A.~Paris, Phys.~Rev.~A {\bf 74}, 012301
  (2006); F.~Buscemi and M.~F.~Sacchi, \emph{ibid.}~{\bf 74}, 052320
  (2006); M.~F.~Sacchi, \emph{ibid.}~{\bf 75}, 012306 (2007).
\bibitem{invert} M.~Ueda and M.~Kitagawa, Phys.~Rev.~Lett. {\bf 68},
  3424 (1992); A.~Imamoglu, Phys.~Rev.~A {\bf 47}, R4577 (1993);
  M.~Ueda, N.~Imoto, and H.~Nagaoka, \emph{ibid.}~{\bf 53}, 3808
  (1996); A.~N.~Koroktov and A.~N.~Jordan, Phys.~Rev.~Lett.~{\bf 97},
  166805 (2006); H.~Terashima and M.~Ueda, Phys.~Rev.~A~{\bf 74},
  012102 (2006).
\bibitem{groene} H.~J.~Groenewold, Int. J.~Theor. Phys.~{\bf 4}, 327
  (1971).
\bibitem{schum-lloyd} B.~Schumacher and M.~A.~Nielsen, Phys.~Rev.~A~{\bf
    54}, 2629 (1996); S.~Lloyd, \emph{ibid.}~{\bf 55}, 1613 (1997).
\bibitem{macca} L.~Maccone, Europhys.~Lett. {\bf 77}, 40002 (2007).
\bibitem{kraus} K.~Kraus, \emph{States, Effects, and Operations:
    Fundamental Notions in Quantum Theory}, Lect.~Notes~Phys.~{\bf
    190}, (Springer-Verlag, 1983).
\bibitem{ozawa-instr} M.~Ozawa, J.~Math.~Phys. {\bf 25}, 79 (1984).
\bibitem{mutual_info} R.~L.~Stratonovich, Prob.~Inf.~Transm.~{\bf 2}, 35
  (1965); C.~Adami and N.~J.~Cerf, Phys.~Rev.~A~{\bf 56}, 3470 (1997).
\bibitem{chi} J.~P.~Gordon, in \emph{Quantum Electronics and Coherent
    Light, Proc. Int. Schoool Phys. ``Enrico Fermi''}, ed. by
  P.~A.~Miles (Academic, New York, 1964); D.~S.~Lebedev and L.~B.~Levitin,
  Inf.~and~Control {\bf 9}, 1 (1966).
\bibitem{holevo} A.~S.~Holevo, Probl.~Inf.~Transm.~{\bf 9}, 110 (1973).
\bibitem{ozawa} M.~Ozawa, J.~Math.~Phys.~{\bf 27}, 759 (1986).
\bibitem{Winter} A.~Winter, Comm.~Math.~Phys.~{\bf 244}, 157 (2004).
\bibitem{Hall} M.~J.~W.~Hall, Phys.~Rev.~A~{\bf 55}, 100 (1997);
  A.~Barchielli and G.~Lupieri, Q.~Inf.~Comp.~{\bf 6}, 16 (2006).
\bibitem{schumacher-westmoreland} B.~Schumacher and M.~D.~Westmoreland,
  Quant.~Inf.~Processing~{\bf 1}, 5 (2002).
\bibitem{schum} B.~Schumacher, Phys.~Rev.~A~{\bf 54}, 2614 (1996).
\bibitem{barnischu} H.~Barnum, M.~A.~Nielsen, and B.~Schumacher,
  Phys.~Rev.~A~{\bf 57}, 4153 (1998).
\bibitem{bus} F.~Buscemi, Phys.~Rev.~Lett.~{\bf 99}, 180501 (2007).
\bibitem{hayashi} M.~Hayashi, \emph{Quantum Information: an
    Introduction} (Springer-Verlag, Berlin Heidelberg, 2006). See
  Eq.~(5.75).
\bibitem{clean} F.~Buscemi \emph{et al.}, J.~Math.~Phys. {\bf 46},
  082109 (2005).

\end{thebibliography}
\end{document}